\documentclass[apjl]{emulateapj}

\newcommand{\rah}{\ensuremath{^{\mathrm{h}}}}
\newcommand{\ram}{\ensuremath{^{\mathrm{m}}}}
\newcommand{\ras}{\ensuremath{^{\mathrm{s}}}}
\newcommand{\peras}{\ensuremath{\mathrm{arcsec}^{-1}}}
\newcommand{\kms}{\mbox{km\ s$^{-1}$}}
\newcommand{\kmsMpc}{\kms~\mbox{Mpc}$^{-1}$}
\newcommand{\Msun}{\ensuremath{M_{\odot}}}
\newcommand{\Msunyr}{\Msun~\mbox{yr}$^{-1}$}
\newcommand{\Lstar}{\ensuremath{L_{*}}}
\newcommand{\RE}{\ensuremath{R_{\mathrm{E}}}}
\newcommand{\aslit}[2]{\mbox{#1\arcsec$\,\times\,$#2\arcsec}}
\newcommand{\slit}[2]{\mbox{#1$\,\times\,$#2}}
\newcommand{\dg}{\ensuremath{^\circ}}
\newcommand{\ten}[1]{\ensuremath{10^{#1}}}
\newcommand{\xten}[2]{\ensuremath{#1\,\times\,10^{#2}}}
\newcommand{\HST}{\textit{HST}}
\newcommand{\Spitzer}{\textit{Spitzer}}

\newcommand{\gband}{\ensuremath{g_{475}}}
\newcommand{\rband}{\ensuremath{r_{625}}}
\newcommand{\iband}{\ensuremath{i_{775}}}
\newcommand{\zband}{\ensuremath{z_{850}}}
\newcommand{\Jband}{\ensuremath{J_{110}}}
\newcommand{\Hband}{\ensuremath{H_{160}}}
\newcommand{\iraci}{[3.6~\micron]}
\newcommand{\iracii}{[4.5~\micron]}
\newcommand{\iraciii}{[5.8~\micron]}
\newcommand{\iraciv}{[8.0~\micron]}

\newcommand{\zmJ}{\ensuremath{\zband - \Jband}}
\newcommand{\JmH}{\ensuremath{\Jband - \Hband}}

\def\figa {
   \begin{figure*}[ht]
   \epsscale{0.93}
   \plotone{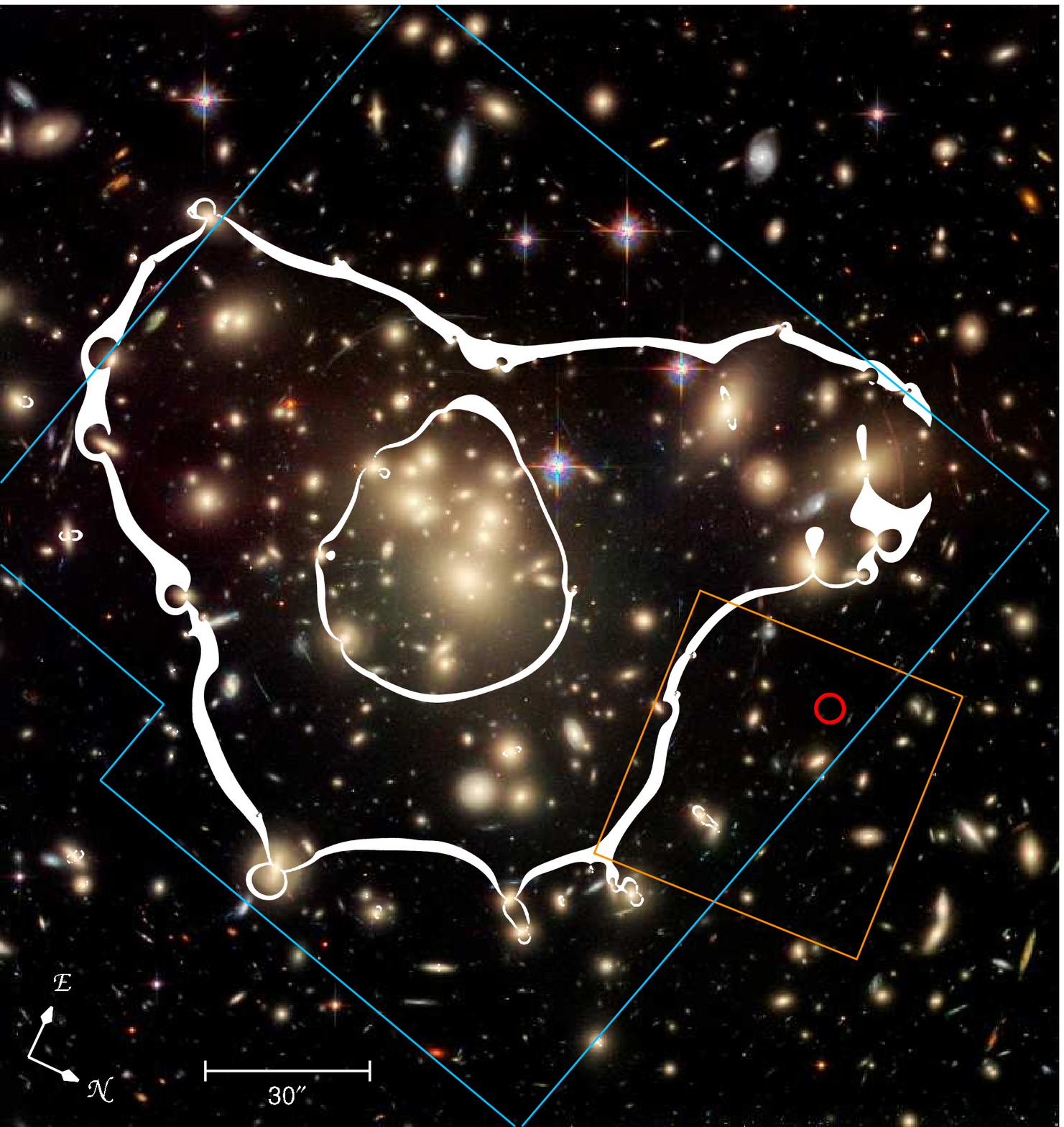}

   \caption{ACS color image (\slit{3.4\arcmin}{3.4\arcmin}) of Abell~1689.  The regions surveyed by our followup NICMOS \Jband\ and \Hband\ images are illustrated by the blue and orange outlines, respectively.  A1689-zD1 is located at J2000 coordinates $\alpha = 13\rah 11\ram 29.96\ras$, $\delta = -1\dg 19\arcmin 18.7\arcsec$ and is denoted by the red circle.  The white contours represent the $z = 7.6$ critical curves ($\mu \ge 200$).}
   \label{fig:a1689}
   \end{figure*}
}

\def\figb {
   \begin{figure*}[ht]
   \epsscale{1.0}
   \plotone{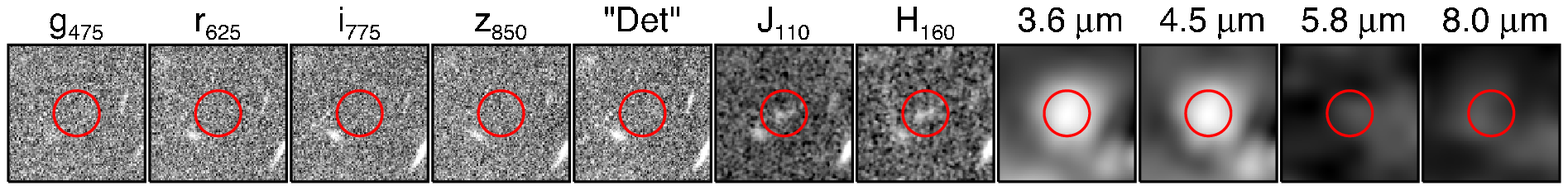}

   \caption{Cutout images (P.A. = 115\dg) from the \HST\ ACS (\gband, \rband, \iband, \zband, and ``Det" -- ``detection"), \HST\ NICMOS (\Jband\ and \Hband), and \Spitzer\ IRAC (3.6, 4.5, 5.8, and 8.0~\micron) data centered on A1689-zD1.  The source is undetected ($< 1\ \sigma$) in the ACS data, including the 20-orbit combined ``detection" image, and also not formally detected at 5.8 and 8.0~\micron.  The fluctuations that are present in the 5.8 and 8.0~\micron\ images near the position of A1689-zD1 are consistent with being noise.  The cutout images are \aslit{5}{5}, corresponding to 25~kpc on a side at $z = 7.6$.}
   \label{fig:cutouts}
   \end{figure*}
}

\def\figc {
   \begin{figure}[t]
   \epsscale{1.0}
   \plotone{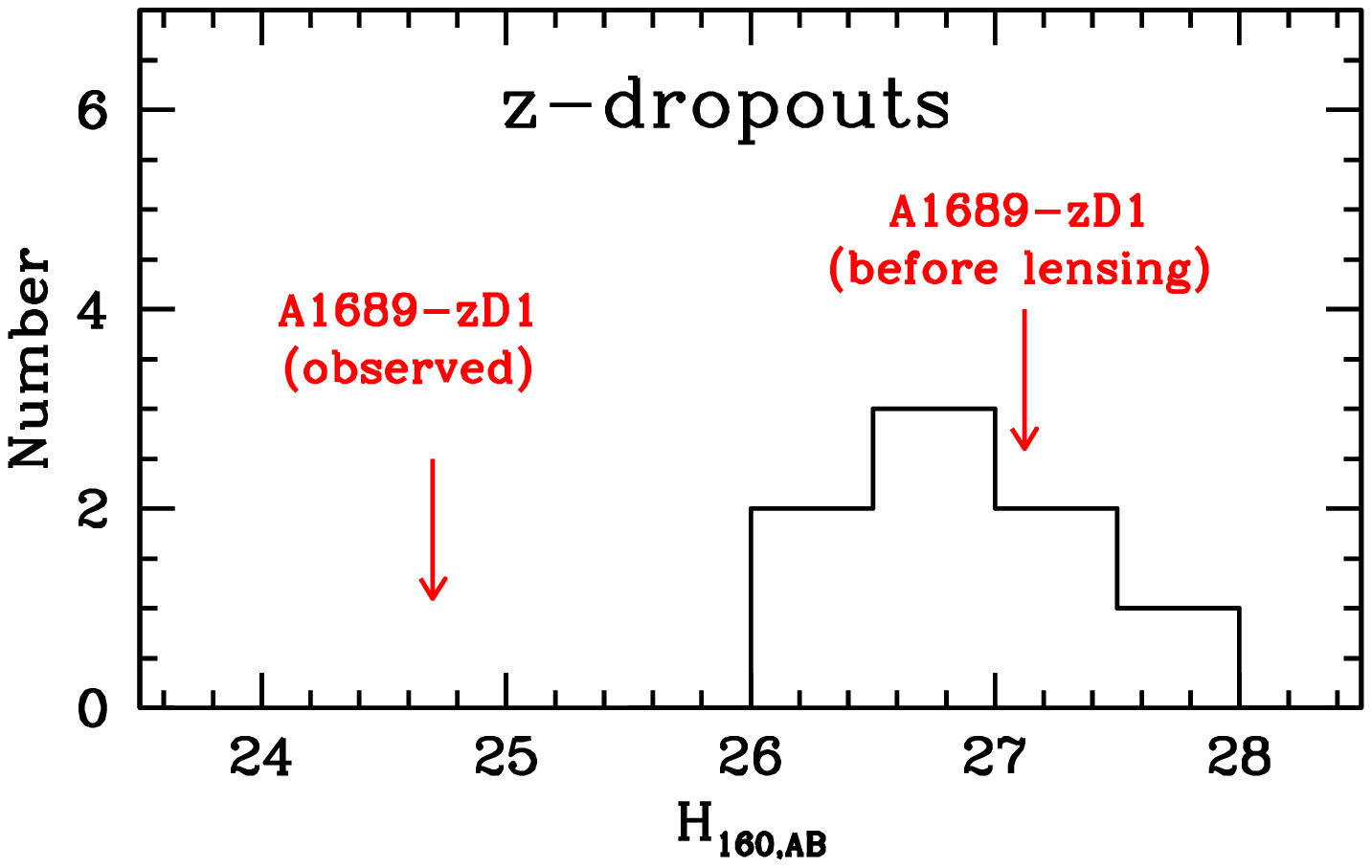}

   \caption{Histogram of \Hband\ magnitudes for \zband-dropout ($z \sim 7-8$) galaxies \citep{Bouwens2008}.  We denote both the observed (uncorrected) and intrinsic (lens-corrected) magnitudes of A1689-zD1.  While the observed brightness of A1689-zD1 is $\sim1.3$ mag brighter than the brightest $z \sim 7-8$ candidate, its intrinsic (lens-corrected) magnitude is very similar to that of other \zband-dropout galaxies.}
   \label{fig:brightness}
   \end{figure}
}

\def\figd {
   \begin{figure}[ht]
   \epsscale{1.00}
   \plotone{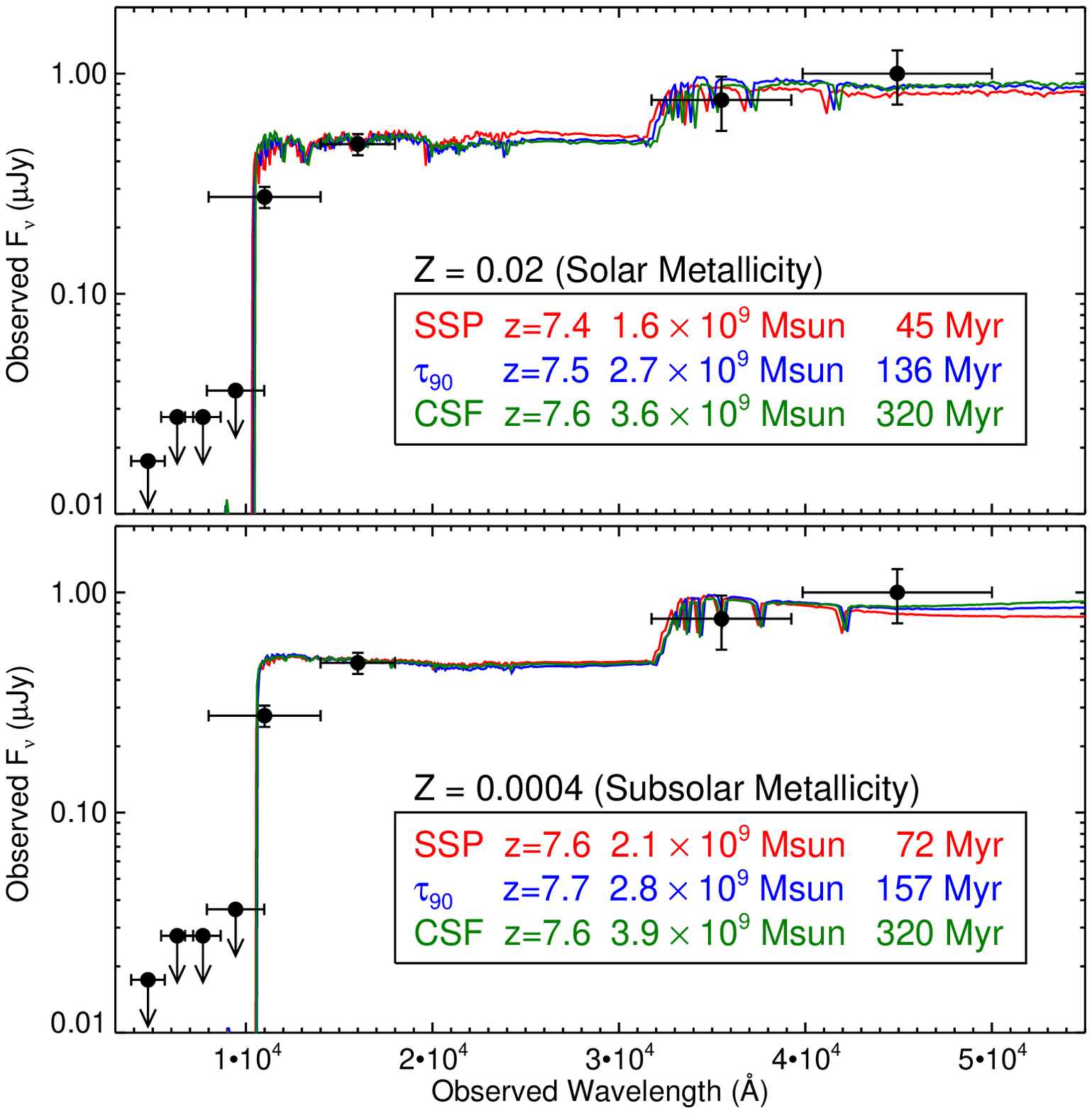}

   \caption{Best-fit stellar population models to the broadband photometry of our $z \sim 7.6$ candidate.  Solar metallicity models are shown in the top panel, with the subsolar ($Z = Z_{\sun} / 50$) models in the bottom panel.  The vertical bars denote the $1\ \sigma$ flux uncertainties and the horizontal bars represent the width of the filter bandpass.  The optical ACS non-detections are shown as $1\ \sigma$ upper limits.}
   \label{fig:models}
   \end{figure}
}

\def\fige {
   \begin{figure}[hb]
   \epsscale{1.0}
   \plotone{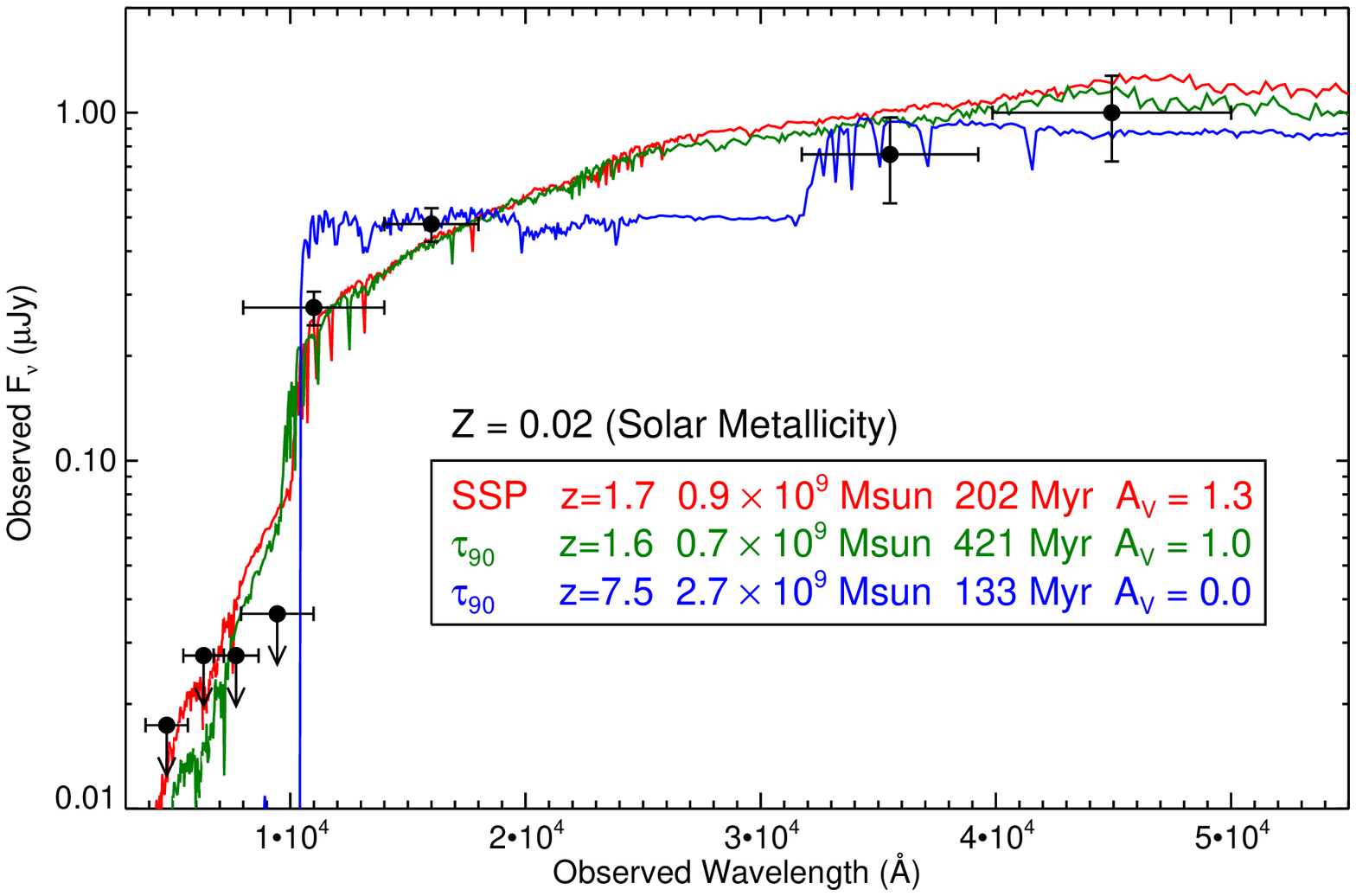}

   \caption{Fits to the observed photometry with stellar population models at $z <4$ ({\em red}, {\em green}) and $z > 4$ ({\em blue}).  Note that the low-redshift SSP ($\chi_{\nu} = 27$) and $\tau_{90}$ ($\chi_{\nu} = 25$) solutions do not agree with the \iband\ and \zband\ upper limits and cannot reproduce the strong \zmJ\ break, even with $\ge 1.0$ mag of extinction.  Also note that because the break in the SED occurs well within the \Jband\ band, the measured flux at 1.1~\micron\ should be appreciably below the continuum level.  We find that the $z \sim 7.5$ fit is quite reasonable ($\chi_{\nu} = 1.1$).}
   \label{fig:lowzsol}
   \end{figure}
}

\def\figf {
   \begin{figure}[hb]
   \epsscale{1.0}
   \plotone{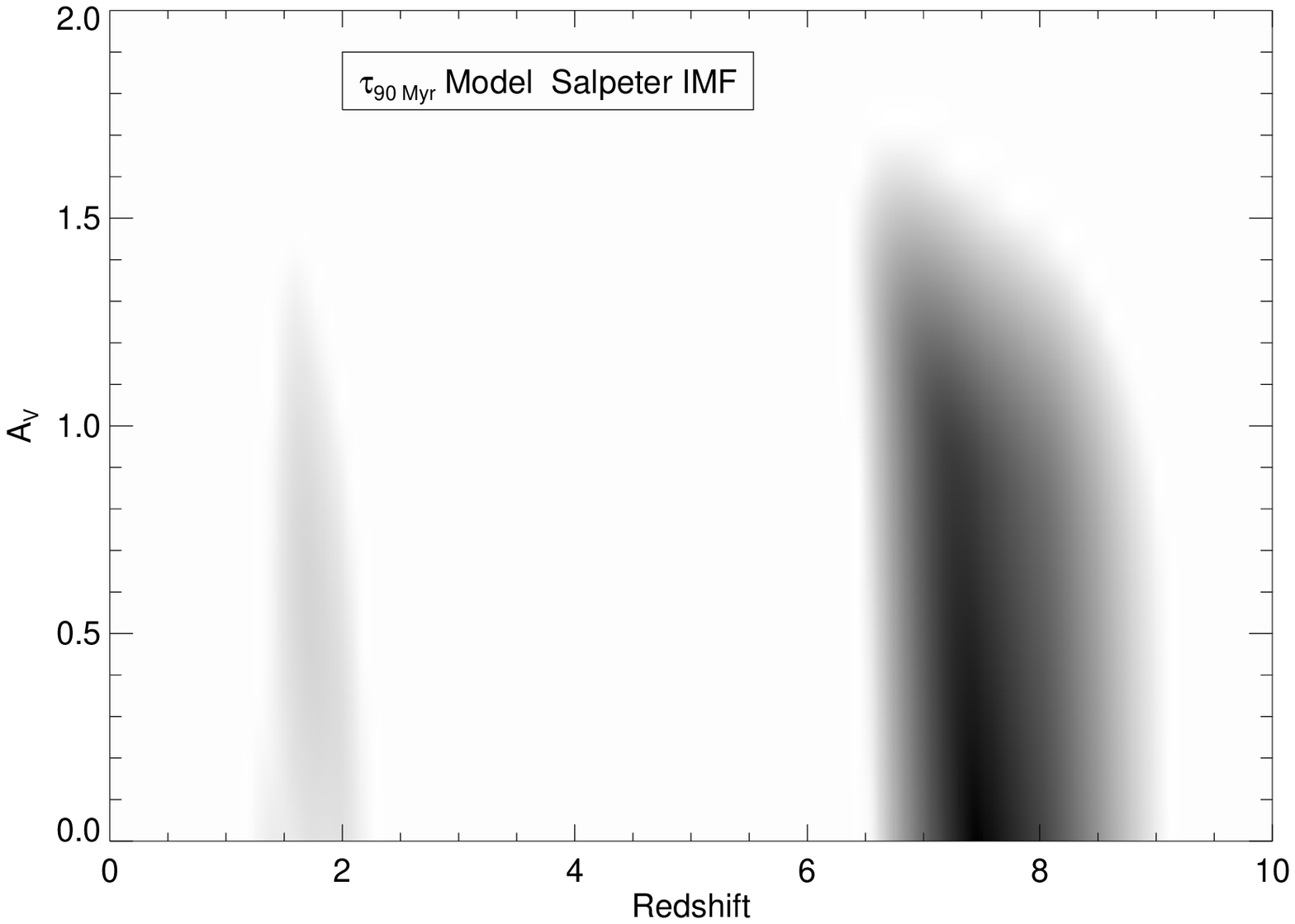}

   \caption{Likelihood plot showing the $\chi^{2}_{\nu}$ values of the best-fitting $\tau_{90}$ solar-metallicity models as a function of redshift and reddening.  Black corresponds to the lowest $\chi^{2}_{\nu}$ value (1.1), the gray contours at $z \sim 1.3 - 2.2$ correspond to $25 \le \chi^{2}_{\nu} < 30$, and white to $\chi^{2}_{\nu} \ge 30$.  The grayscale colors are displayed with a square-root stretch to compress the dynamic range of $\chi^{2}_{\nu}$ values.  Clearly, the best low-redshift solutions ($1.5 \la z \la 2.0$) provide a much poorer fit ($\chi^{2}_{\nu} \ge 25$) to the data than the high-redshift ($z > 6.5$) solutions.}
   \label{fig:likelihood}
   \end{figure}
}

\def\figg {
   \begin{figure}[ht]
   \epsscale{1.0}
   \plotone{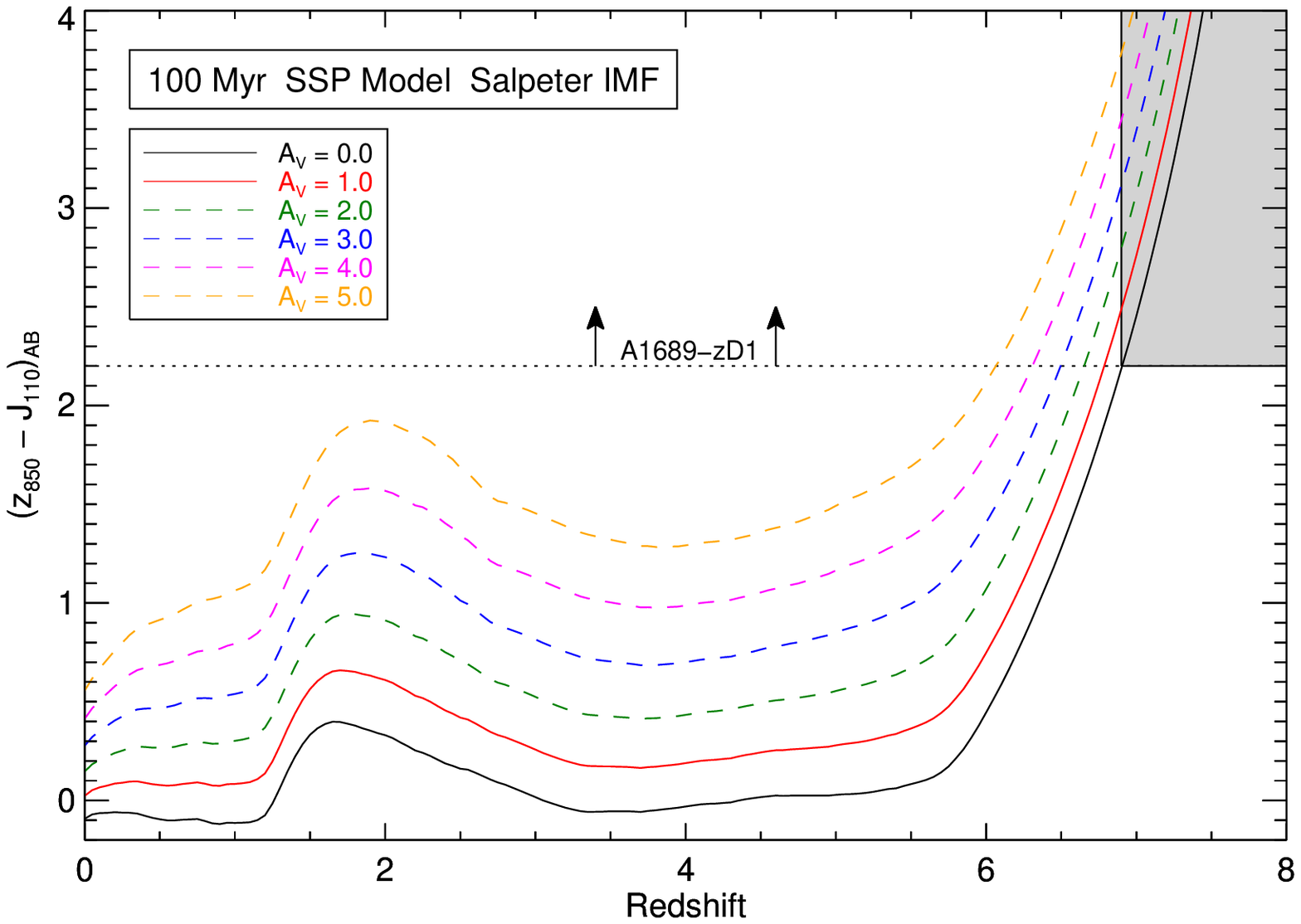}

   \caption{Plot showing (\zmJ)$_{\mbox{\scriptsize{AB}}}$ color as a function of redshift for a range of reddened ($0 \le A_{V} \le 5$) 100~Myr SSP solar metallicity models.  Even with extremely large values of the reddening, the low-redshift models are unable to reproduce the strong \zmJ\ break of 2.2 mag observed for A1689-zD1.  Because we must fit the entire SED of the galaxy, the reddening cannot be increased arbitrarily to fit only the \zmJ\ break.  For example, reddening values of $A_{V} \ga 1.5$ ({\em dashed lines}) produce fits that are inconsistent with the observed \Hband$-$\iracii\ color of A1689-zD1.  The gray box shows the redshift range where the stellar population models are consistent with the \zmJ\ color.}
   \label{fig:zmjredshift}
   \end{figure}
}

\def\figh {
   \begin{figure*}[ht]
   \epsscale{1.0}
   \plottwo{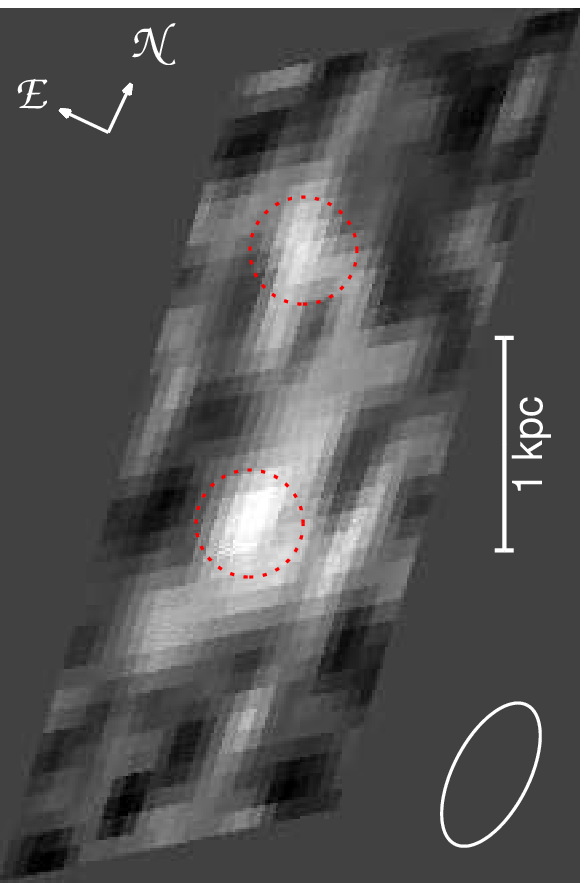}{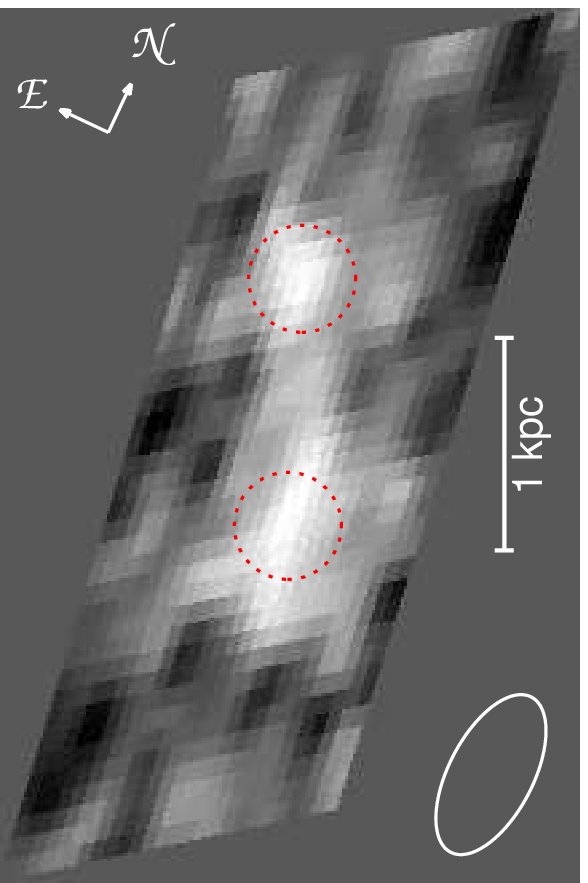}

   \caption{Source-plane deprojection of the NICMOS \Jband\ ({\em left}) and \Hband\ ({\em right}) images of A1689-zD1.  The morphology is remarkably consistent in both images, showing an extended morphology that spans $\sim0.4\arcsec$ (2.0~kpc) in the source plane.  A1689-zD1 appears to be comprised of two compact knots, denoted by the dotted red circles, connected by a lower-luminosity region.  Each knot appears to have an intrinsic half-light radius that is less than 300~pc (see \S~\ref{sect:morphology}).  The solid white ellipses represent the NIC3 \Jband\ and \Hband\ PSFs (0.33\arcsec\ and 0.37\arcsec\ FWHM, respectively), delensed to the source plane at $z = 7.6$.}
   \label{fig:srcplane}
   \end{figure*}
}

\def\taba {
\tabletypesize{\tiny}
\begin{deluxetable*}{lcccccccccc}
\tablecolumns{11}
\tablewidth{0pt}
\tablecaption{Photometry Summary}

\tablehead{\colhead{Measurement} & \colhead{\gband} & \colhead{\rband} & \colhead{\iband} & \colhead{\zband} & \colhead{\Jband} & \colhead{\Hband} & \colhead{\iraci} & \colhead{\iracii} & \colhead{\iraciii} & \colhead{\iraciv} }

\startdata
A1689-zD1 Observed Magnitude   & $> 28.3$   & $> 27.8$   & $> 27.8$   & $> 27.5$   & 25.3 (0.1) & 24.7 (0.1) & 24.2 (0.3) & 23.9 (0.3) & $> 23.9$ & $> 23.4$ \\
A1689-zD1 Intrinsic Magnitude  & $> 30.7$   & $> 30.2$   & $> 30.2$   & $> 29.9$   & 27.7 (0.1) & 27.1 (0.1) & 26.6 (0.3) & 26.3 (0.3) & $> 26.3$ & $> 25.8$ \\
Companion Magnitude\tablenotemark{a}     & 27.4 (0.1) & 26.4 (0.1) & 26.1 (0.1) & 25.9 (0.1) & 25.0 (0.1) & 24.3 (0.1) & 25.0 (0.3) & 24.3 (0.3) & $> 23.9$ & $> 23.4$ \\
\enddata

\tablecomments{Errors ($1\ \sigma$) are given in parenthesis.  Intrinsic values were calculated assuming a cluster magnification of $\mu = 9.3$ (see \S~\ref{sect:srcmag}).  The \iraciii\ and \iraciv\ magnitudes are 1 and $2\ \sigma$ upper limits, respectively.  Because Abell~1689 lies in the plane of the ecliptic, the zodiacal light background is high and increases in the \iraciii\ and \iraciv\ bands, limiting our ability to detect the object in these two bands.}
\tablenotetext{a}{We include the photometry of this nearby source because it overlaps A1689-zD1 in the IRAC images and may therefore contaminate mildly our flux measurement of A1689-zD1.}
\label{tbl:photometry}
\end{deluxetable*}
}

\def\tabb {
\tabletypesize{\footnotesize}
\begin{deluxetable*}{cccccccc}
\tablecolumns{8}
\tablewidth{0pt}
\tablecaption{Best-fit Stellar Population Model Results}

\tablehead{
                     &                   &                                    &  Mass\tablenotemark{c} &  Age\tablenotemark{d} &  SFR      &       &                \\
SFH\tablenotemark{a} & $z_{\mbox{\scriptsize phot}}$ & $z_{\mbox{\scriptsize form}}$\tablenotemark{b} & (\ten{9} \Msun)        & (Myr)                 & (\Msunyr) & $A_V$ & $\chi^{2}_{\nu}$
}

\startdata
\cutinhead{\sc Solar Metallicity ($Z = 0.02$)}

         SSP & $7.4 \pm 0.3$ &  7.8 & $1.6 \pm 0.7$  &  45 &   0 & $0.0 \pm 0.2$ & 1.4 \\
$\tau_{90}$  & $7.5 \pm 0.3$ &  9.7 & $2.7 \pm 0.4$  & 136 & 3.2 & $0.0 \pm 0.2$ & 1.2 \\
  CSF        & $7.6 \pm 0.4$ & 63.4 & $3.6 \pm 1.4$  & 320 & 6.9 & $0.1 \pm 0.2$ & 0.9 \\

\cutinhead{\sc Subsolar Metallicity ($Z = 0.0004$)}
         SSP & $7.6 \pm 0.4$ &  8.3 & $2.1 \pm 0.9$  &  72 &   0 & $0.0 \pm 0.2$ & 1.1 \\
$\tau_{90}$  & $7.7 \pm 0.3$ & 10.5 & $2.8 \pm 1.1$  & 157 & 2.4 & $0.0 \pm 0.2$ & 0.7 \\
  CSF        & $7.6 \pm 0.4$ & 63.4 & $3.9 \pm 1.6$  & 320 & 7.6 & $0.3 \pm 0.2$ & 0.6 \\
\enddata

\tablenotetext{a}{Star-formation history (SFH):  simple (single-burst) stellar population (SSP), exponentially declining SFR ($\tau$ models) where the subscript represents the $e$-folding time in Myr, or constant SFR (CSF) models.}
\tablenotetext{b}{Formation redshift calculated from the fitted redshift and (unweighted) stellar age.}
\tablenotetext{c}{Best-fit stellar mass, corrected by the magnification at the fitted $z_{\mbox{\scriptsize phot}}$ redshift.}
\tablenotetext{d}{SFR-weighted mean stellar age \citep[cf.][]{Forster2004}.}
\label{tbl:models}
\end{deluxetable*}
}

\journalinfo{The Astrophysical Journal, 678:647--654, 2008 May 10}
\slugcomment{Received 2007 September 21; accepted 2008 January 8}
\shorttitle{STRONGLY LENSED LYMAN BREAK GALAXY CANDIDATE}
\shortauthors{BRADLEY ET AL.}

\begin{document}

\title{Discovery of a Very Bright Strongly Lensed Galaxy Candidate at
\lowercase{$z \approx 7.6}$\altaffilmark{1}}

\author{L.D.~Bradley\altaffilmark{2}, R.J.~Bouwens\altaffilmark{3},
H.C.~Ford\altaffilmark{2}, G.D.~Illingworth\altaffilmark{3},
M.J.~Jee\altaffilmark{4}, N.~Ben\'{\i}tez\altaffilmark{5},
T.J.~Broadhurst\altaffilmark{6}, M.~Franx\altaffilmark{7},
B.L.~Frye\altaffilmark{8}, L.~Infante\altaffilmark{9},
V.~Motta\altaffilmark{10}, P.~Rosati\altaffilmark{11},
R.L.~White\altaffilmark{12}, W.~Zheng\altaffilmark{2}}

\altaffiltext{1}{Based on observations made with the NASA/ESA {\em
Hubble Space Telescope}, obtained at the Space Telescope Science
Institute, which is operated by the Association of Universities
for Research in Astronomy under NASA contract NAS5-26555.  Based on
observations made with the {\em Spitzer Space Telescope}, which is
operated by the Jet Propulsion Laboratory, California Institute of
Technology under NASA contract 1407.}

\altaffiltext{2}{Department of Physics and Astronomy, Johns Hopkins
University, 3400 North Charles Street, Baltimore, MD 21218.}

\altaffiltext{3}{UCO/Lick Observatory, Department of Astronomy
and Astrophysics, University of California Santa Cruz, Santa Cruz,
CA 95064.}

\altaffiltext{4}{Department of Physics, 1 Shields Avenue, University
of California, Davis, CA 95616.}

\altaffiltext{5}{Instituto de Matem\'{a}ticas y F\'{\i}sica
Fundamental (CSIC), C/Serrano 113-bis, 28006, Madrid, Spain}

\altaffiltext{6}{School of Physics and Astronomy, Tel Aviv
University, Tel Aviv 69978, Israel.}

\altaffiltext{7}{Leiden Observatory, Leiden University, Postbus 9513,
2300 RA Leiden, Netherlands.}

\altaffiltext{8}{Department of Physical Sciences, Dublin City
University, Dublin 9, Ireland.}

\altaffiltext{9}{Departmento de Astronom\'{\i}a y Astrof\'{\i}sica,
Pontificia Universidad Cat\'{o}lica de Chile, Casilla 306, Santiago
22, Chile.}

\altaffiltext{10}{Departamento de F\'{\i}sica y Astronom\'{\i}a,
Universidad de Valpara\'{\i}so, Av. Gran Breta\~na 1111,
Valpara\'{\i}so, Chile.}

\altaffiltext{11}{European Southern Observatory,
Karl-Schwarzschild-Strasse 2, D-85748 Garching, Germany.}

\altaffiltext{12}{Space Telescope Science Institute, 3700 San Martin
Drive, Baltimore, MD 21218.}

\begin{abstract}

Using {\em Hubble Space Telescope (HST)} and {\em Spitzer} IRAC
imaging, we report the discovery of a very bright strongly lensed
Lyman break galaxy (LBG) candidate at $z\sim7.6$ in the field of the
massive galaxy cluster Abell 1689 ($z=0.18$).  The galaxy candidate,
which we refer to as A1689-zD1, shows a strong $z_{850}-J_{110}$
break of at least 2.2 mag and is completely undetected ($< 1\
\sigma$) in {\em HST} Advanced Camera for Surveys (ACS) $g_{475}$,
$r_{625}$, $i_{775}$, and $z_{850}$ data.  These properties,
combined with the very blue $J_{110}-H_{160}$ and $H_{160}-[4.5
\mu m]$ colors, are exactly the properties of an $z\sim7.6$ LBG
and can only be reasonably fit by a star-forming galaxy at $z=7.6
\pm 0.4$ ($\chi^{2}_{\nu} = 1.1$).  Attempts to reproduce these
properties with a model galaxy at $z<4$ yield particularly poor
fits ($\chi^{2}_{\nu} \ge 25$).  A1689-zD1 has an observed (lensed)
magnitude of 24.7 AB ($8\ \sigma$) in the NICMOS $H_{160}$ band and is
$\sim1.3$ mag brighter than the brightest-known $z_{850}$-dropout
galaxy.  When corrected for the cluster magnification of $\sim9.3$
at $z\sim7.6$, the candidate has an intrinsic magnitude of
$H_{160}=27.1$ AB, or about an $L_{*}$ galaxy at $z\sim7.6$.
The source-plane deprojection shows that the star formation is
occurring in compact knots of size $\la 300$ pc.  The best-fit
stellar population synthesis models yield a median redshift of 7.6,
stellar masses $(1.6-3.9) \times 10^{9} M_{\sun}$, stellar ages
$45-320$ Myr, star-formation rates $\la7.6 M_{\sun}$ yr$^{-1}$, and
low reddening with $A_V \le 0.3$.  These properties are generally
similar to those of LBGs found at $z\sim5-6$.  The inferred stellar
ages suggest a formation redshift of $z\sim8-10$ ($t\la0.63$ Gyr).
A1689-zD1 is the brightest observed, highly reliable $z>7.0$ galaxy
candidate found to date.
\end{abstract}


\section{Introduction}

One of the most important frontiers of observational cosmology
is the characterization of the earliest galaxies in the universe.
The {\em Hubble Space Telescope} (\HST) has been at the forefront of
such high-redshift searches, which have recently provided significant
insights to the mass assembly and buildup of the earliest galaxies
($z \ga 6$, $t \la 0.95$~Gyr) and the contribution of star formation
to cosmic reionization \citep{Lehnert2003, Bunker2004, Yan2004a,
Bouwens2006}.  Recent {\em WMAP} optical depth measurements indicate
that reionization began between $z \sim 8 - 15$ \citep{Spergel2007,
Page2007}, while measurements of Gunn-Peterson absorption troughs
in SDSS quasars suggest reionization was complete by $z \sim 6$
\citep{Fan2006}.  High-redshift galaxy searches are now probing the
era of cosmic reionization \citep[][but also see \citealt{Chary2007}
regarding the \citealt{Mobasher2005} object]{Hu2002, Kneib2004,
Bouwens2004, Egami2005, Mobasher2005, Stern2005, Richard2006,
Iye2006, Bouwens2006Nature, Stark2007, Bouwens2008} and are beginning
to characterize the luminosity density and star-formation history of
this important epoch \citep{Yan2004a, Bunker2004, Giavalisco2004b,
Yan2005, Stanway2005, Bouwens2006Nature, Bouwens2006}.  Early
indications suggest that low-luminosity star-forming galaxies at high
redshift ($z > 6$) likely play a significant role in reionizing the
universe \citep{Lehnert2003, Yan2004a, Bouwens2006}, but this needs
to be verified by studying fainter galaxies at higher redshifts;
galaxies at $z \ga 7$ represent the current high-redshift frontier.

\figa

Strong gravitational lenses produced by massive galaxy clusters
provide an opportunity to observe the high-redshift universe
in unprecedented detail.  The large magnifications provided by
nature's ``cosmic telescopes'' increase both the flux and size of
background sources.  These gains make it possible to detect faint
high-redshift galaxies \citep{Kneib1996, Pello1999, Ellis2001,
Hu2002, Kneib2004, Egami2005, Frye2007} without requiring a huge
investment of observing time, such as that dedicated to the Hubble
Ultra Deep Field \citep{Beckwith2006}.  The key to making the
subsequent analyses viable is having detailed cluster magnification
maps that constrain the ``optics'' of the ``cosmic telescope''.
Our team has been among the first to overcome this difficulty
by using multiband \HST\ ACS data to perform detailed studies of
massive rich galaxy clusters, including Abell~1689 \citep[][Coe et
al. 2008, in preparation]{Broadhurst2005, Zekser2006} and CL0024+17
\citep{Jee2007}, to a precision useful for studying the properties
of the background galaxy population.

In this paper, we present the discovery of a bright strongly
lensed Lyman break galaxy (LBG) candidate at $z \sim 7.6$ in the
field of the massive cluster Abell~1689.  We adopt a cosmology
with $\Omega_{m} = 0.3$, $\Omega_{\Lambda} = 0.7$, and $H_{0} =
70$~\kmsMpc.  This provides an angular scale of 5.0~kpc~\peras\
at $z = 7.6$.  All magnitudes are given in the AB photometric
system \citep{Oke1974}.

\taba
\figb

\section{Observations and Photometry}

\subsection{\HST\ ACS and NICMOS Data}

Abell~1689 is a well-studied and massive galaxy cluster at $z =
0.183$.  With an unequaled Einstein radius of $\sim$50\arcsec, it
is the strongest gravitational lens known \citep{Broadhurst2005}.
We observed the central 3.4\arcmin\ $\times$ 3.4\arcmin\ of
Abell~1689 with a single pointing of the ACS Wide Field Camera
(WFC) in 2002 June (see Fig.~\ref{fig:a1689}).  The observations
consisted of 20 orbits divided among four broadband filters:
F475W (\gband; 9500~s), F625W (\rband; 9500~s), F775W (\iband;
11800~s), and F850LP (\zband; 16600~s).  In 2005 May, we followed
up our ACS observations with 18 orbits of Near Infrared Camera and
Multi-Object Spectrometer (NICMOS)/NIC3 F110W (\Jband) imaging to
search for \zband-dropout candidates.  The observations cover the
central 2.5\arcmin\ $\times$ 2.5\arcmin\ high-magnification region
of the cluster with a 9-pointing 3 $\times$ 3 mosaic (5376~s per
NIC3 field).

The bright LBG candidate, which we refer to as A1689-zD1, was
discovered by comparing the ACS/WFC \zband\ and NICMOS/NIC3 \Jband\
data.  The galaxy has an observed magnitude of $25.3 \pm 0.1$ in
the NICMOS \Jband\ band and a strong \zmJ\ break of $> 2.2$ mag (see
Table~\ref{tbl:photometry}).  With a magnification of $\mu \approx
9.3$, A1689-zD1 is strongly lensed by the foreground cluster and
is $\sim$1.3 mag brighter (as observed) than the brightest-known
\zband-dropout ($z \sim 7-8$) galaxy \citep{Bouwens2006Nature,
Bouwens2008}.

A1689-zD1 is completely undetected ($< 1\ \sigma$) in the ACS
\gband, \rband, \iband, and \zband\ data as well as in the 20-orbit
ACS rms-weighted ``detection" image constructed by co-adding all
of the ACS images (see Fig.~\ref{fig:cutouts}).  The $1\ \sigma$
detection limits for this galaxy are 28.3, 27.8, 27.8, and 27.5 in
the \gband, \rband, \iband, and \zband\ bands, respectively.

To verify that A1689-zD1 was not a low-redshift reddened interloper,
we acquired a single-orbit NIC3 F160W (\Hband) image of the
source in 2007 June.  Combining these data with the earlier
NICMOS \Jband\ data, we measure a \JmH\ color of $0.6 \pm 0.2$.
This is exactly what one would expect for an LBG at $z \sim 7.4
- 7.7$ \cite[see, e.g., Fig.~1 in the Supplementary Information
of][]{Bouwens2006Nature}, but much bluer than a reddened object at
lower redshift.  Note that the strong \zmJ\ break and the blue color
redward of the break are precisely the two criteria used to define
high-redshift LBGs.  Together this information provides compelling
evidence that our source is a star-forming galaxy at $z \sim 7.4
- 7.7$.

A1689-zD1 is not detected in the individual \Jband\ and \Hband\
dither exposures ($\sim$640~sec, 4 dithers per orbit), but there
also does not appear to be any artifacts (e.g., cosmic rays) at the
object position.  Given the similar morphology in both the \Jband\
and \Hband\ images, which were taken about two years apart, and
also that A1689-zD1 was located at different detector positions in
the \Jband\ and \Hband\ images, we are confident that A1689-zD1 is
not a spurious detection.

We present in Figure~\ref{fig:brightness} the observed and intrinsic
(lens-corrected) \Hband\ magnitudes of A1689-zD1 relative to other $z
\sim 7-8$ galaxy candidates \citep{Bouwens2008}.  While the observed
(uncorrected) brightness of A1689-zD1 is $\sim$1.3 mag brighter than
the brightest $z \sim 7-8$ candidate, its intrinsic (lens-corrected)
magnitude is very similar to that of other \zband-dropout galaxies.

\subsection{\Spitzer\ IRAC Data}

Further evidence supporting the interpretation of A1689-zD1 as
a $z \sim 7.6$ LBG comes from archival \Spitzer\ IRAC data (GO
20439, PI: Egami).  In 2006, ultradeep IRAC imaging was obtained
for Abell~1689 in the 3.6, 4.5, 5.8, and 8.0~\micron\ bands (11.1
hours per band).  Our LBG candidate is detected as a point-source
in the IRAC 3.6 and 4.5~\micron\ bands.

The large size of the IRAC point-spread function (PSF; FWHM
$\sim2$\arcsec) poses a difficulty in that neighboring objects may
significantly contaminate the photometry.  In our case, we find
that A1689-zD1 is blended with a nearby foreground object, which is
separated by $\sim 1.5\arcsec$.  As seen in Figure~\ref{fig:cutouts},
the foreground galaxy is rather blue and we estimate its photometric
redshift \citep{Benitez2000} is $\sim2.2$.  To correct for the
contamination of nearby objects, we convolved the NICMOS \Hband\
detections of A1689-zD1 and its neighbors with the IRAC PSF
for a given band.  The \Hband\ PSF-convolved objects were then
simultaneously fitted to the IRAC data, allowing the relative flux
scalings as free parameters.  The fitted neighbors, which contributed
$\sim32$\% to the original flux, were then subtracted from the
IRAC data, leaving the LBG source.  We estimate that the neighbor
subtraction induces an additional $\sim$0.2 mag of uncertainty to
the IRAC photometry.

To minimize background noise and to further limit flux contamination
from nearby objects, we performed the \Spitzer\ IRAC photometry
using a 2.5\arcsec\ diameter aperture.  Assuming a stellar profile
to correct the fluxes for light falling outside the aperture,
we applied aperture corrections of 0.56 and 0.60 mag to the
[3.6 \micron] and [4.5 \micron] bands, respectively.  We measure
observed magnitudes of 24.2 and 23.9 in the [3.6 \micron] and
[4.5 \micron] bands, respectively.  The [5.8 \micron] and [8.0
\micron] upper limits are 23.9 ($1\ \sigma$) and 23.4 ($2\ \sigma$),
respectively.  The broadband photometry of A1689-zD1 is summarized
in Table~\ref{tbl:photometry} and the \HST/ACS, \HST/NICMOS, and
\Spitzer\ IRAC cutout images are shown in Figure~\ref{fig:cutouts}.

\figc

\section{Source Magnification}
\label{sect:srcmag}

It is valuable to calculate the intrinsic properties of A1689-zD1 by
using strong lensing models available for Abell~1689 \citep[e.g,][Coe
et al. 2008, in preparation]{Broadhurst2005}.  For an improved
version of the \cite{Broadhurst2005} strong lensing model and a
redshift of $z \sim 7.6$ (see \S~\ref{sect:models}), we estimate
a magnification $\mu$ of $\sim 9.3$, which appears to be roughly
consistent with other models in the literature \citep[][Coe et al.\
2008, in preparation]{Halkola2006, Limousin2007}.  The improved
Broadhurst lensing model is based on 42 sets of multiple images (12
more than \citealt{Broadhurst2005}), with several new counterimages
close to the cluster center.  Based on the mean slope of the radial
mass profile, which is the most influential model parameter, we
estimate that the $1\ \sigma$ error on the magnification is at least
15\%.  The error is derived from the allowed range in radial slope
parameters that are consistent with the data.  However, comparison
with other Abell~1689 models, e.g., Coe et al. 2008 (in preparation;
$\mu = 7.3$ for A1689-zD1), suggests that a more realistic value of
the systematic uncertainty in our magnification value is $\sim$25\%.
When corrected for the cluster magnification of 9.3, the intrinsic
magnitudes of A1689-zD1 are 27.7, 27.1, 26.6, and 26.3 in the \Jband,
\Hband, \iraci, and \iracii\ bands, respectively.

\tabb
\figd

\section{Stellar Population Models}
\label{sect:models}

To constrain the stellar populations of A1689-zD1, we fit the
stellar population models of \cite{Bruzual2003} to the multiband
photometry.  We used a \cite{Salpeter1955} initial mass function
(IMF) with mass cutoffs of 0.1 and 100 \Msun\ and explored models
with both solar ($Z = 0.02$) and subsolar ($Z = 0.0004 = Z_{\sun}
/ 50$) metallicities.  The effect of dust reddening is included
in the models using the \cite{Calzetti2000} obscuration law.
We correct for Lyman-series line-blanketing and photoelectric
absorption following the prescription of \cite{Madau1995}.  In the
stellar population model fits, we constrain the stellar age to be
less than the age of the universe at the fit redshift (e.g., 0.7 Gyr
at $z \sim 7.5$).  We considered several star-formation histories
(SFH) including a simple (single-burst) stellar population (SSP),
an exponentially declining star-formation rate (SFR; $\tau$ models)
with $e$-folding times between 10 and 100 Myr, and constant SFR
(CSF) models with SFRs up to 30~\Msunyr.

The best-fit stellar population models are shown in
Figure~\ref{fig:models} and the parameters are given in
Table~\ref{tbl:models}.  For both solar and subsolar metallicities,
we find acceptable fits for the models.  The best-fit redshifts are
in the range from $7.4 < z < 7.7$ with an $1\ \sigma$ error of $0.3 -
0.4$ and a median value of 7.6.  A crosscheck of the redshift was
also performed using the Bayesian photometric redshift (BPZ) code
of \cite{Benitez2000} and showed that A1689-zD1 is certainly $z >
7$, with the probably concentrated at $7 < z < 8$ and a spectral
type between a \cite{Kinney1996} SB2 and SB3.

To examine the possibility that this source could be a
highly-reddened galaxy at low redshift, we re-fit the stellar
population models by restricting the redshift to be $z \le 4$.
The best-fit $\tau_{90}$ low-redshift solution is at a redshift
of $z = 1.6$ with $A_V = 1.0$, but the fit is particularly poor
($\chi_{\nu}$ = 25).  As seen in Figure~\ref{fig:lowzsol},
the low-redshift models do not agree with the upper limits
on the \iband\ and \zband\ fluxes and do not reproduce the
strong \zmJ\ break, even with $\ge 1.0$ mag of extinction.
In Figure~\ref{fig:likelihood}, we plot the $\chi^{2}_{\nu}$
values of the best-fitting $\tau_{90}$ solar-metallicity models as
a function of redshift and reddening.  We find that the low-redshift
solutions ($1.5 \la z \la 2.0$) all provide a much poorer fit to the
data than the high-redshift ($z > 6.5$) solutions.  To further make
the point, we looked at a single observational parameter, the \zmJ\
color.  In Figure~\ref{fig:zmjredshift}, we plot the \zmJ\ color as
a function of redshift for a range of reddened ($0 \le A_{V} \le 5$)
100~Myr SSP solar metallicity models.  The low-redshift models, even
with extremely large values of the reddening, are unable to reproduce
the strong \zmJ\ break of 2.2 mag observed for A1689-zD1.

\fige

For our preferred high-redshift solutions, we find stellar masses
of $\xten{(1.6 - 3.6)}{9}~\Msun$ for the solar metallicity models
and slightly higher masses of $\xten{(2.1 - 3.9)}{9}~\Msun$ for the
subsolar metallicity models.  In a similar fashion, the subsolar
metallicity models generally produce slightly older ages ($72 -
320$ Myr) than the solar metallicity models ($45 - 320$ Myr).
These stellar mass and age estimates are in the range of those
previously estimated by \cite{Labbe2006} for a HUDF-selected
\zband-dropout ($z \sim 7$) sample.  The derived instantaneous SFRs
show modest star formation with rates up to 7.6~\Msunyr.  Most of
the models require little reddening, with $A_{V} \le 0.3$, which
is in agreement with results for $z \sim 6$ LBGs \citep{Yan2005,
DowHygelund2005, Eyles2006}.   While we assumed a Salpeter IMF,
fitting models using a \cite{Chabrier2003} IMF yields $\sim 1.5$
times lower stellar masses and roughly similar ages.

\figf
\figg
\figh

\section{Source-Plane Deprojection}

\subsection{Morphology}
\label{sect:morphology}

The detailed Abell~1689 cluster deflection map (see
\S~\ref{sect:srcmag}) allows us to deproject the NICMOS \Jband\
and \Hband\ images of A1689-zD1 to the source plane at $z \sim 7.6$.
The substantial magnification of the source provides us with a unique
opportunity to examine the morphology of a $z \sim 7.6$ galaxy at
very high resolution.  The deprojected images of A1689-zD1 in the
\Jband\ and \Hband\ bands are presented in Figure~\ref{fig:srcplane}.
The galaxy shows an extended morphology in both the image and source
planes, spanning $\sim0.4\arcsec$ (2.0~kpc) in the source plane.
The remarkable consistency of the morphology in both the \Jband\
and \Hband\ bands as well as the $> 8\ \sigma$ detection in each
band provides strong evidence that the source is real and not a
spurious detection.

In both images, A1689-zD1 appears to be comprised of at least two
bright, marginally resolved knots connected by a lower-luminosity
region.  The brighter knot is located to the southwest of the
secondary knot and is separated from it by $\sim0.26\arcsec$
(1.3~kpc).  Performing fits to the individual knots with GALFIT
\citep{Peng2002}, we find that most of the luminosity of A1689-zD1
is emitted from the knots.  The brighter (fainter) knot contributes
$\sim$ 2/3 ($\sim$ 1/3) of the total luminosity of A1689-zD1.
The \JmH\ colors of the two knots are roughly consistent, with values
of $0.5 \pm 0.2$ and $0.8 \pm 0.2$ for the brighter and fainter
knots, respectively.  The knots are compact and have half-light
radii $\la 0.06\arcsec$ ($0.3$~kpc) in the source plane.

The concentration of a significant fraction of the total flux in
compact star-forming knots is consistent with findings for starburst
galaxies over a wide range of redshifts \citep[e.g.,][]{Meurer1995,
Law2007, Smail2007, Overzier2007} and as shown quite dramatically in
the $z = 4.92$ lensed galaxy pair in CL~1358+62 \citep{Franx1997}.
The knots in A1689-zD1 are likely separate star-forming regions
within the galaxy, but they could conceivably be interpreted as
small star-forming galaxies merging at high redshift.

\subsection{Half-Light Radius}

We calculated the half-light radius of A1689-zD1 by defining
the circular aperture containing half the total Kron flux
\citep{Kron1980}.  For A1689-zD1 we measure a half-light radius of
0.48\arcsec, which translates to $\sim 0.19\arcsec$ (0.96~kpc) in
the source-plane at $z \sim 7.6$.  The size translation was derived
by comparing the size of the object along its major axis in both
the image and source planes.  As demonstrated by the PSF ellipses
in Figure~\ref{fig:srcplane}, the magnification of A1689-zD1 is
greater along its minor axis than its major axis.  This results in
a source-to-image-plane size scaling along the galaxy major axis
of 2.5, which is slightly smaller than the value one would obtain
by assuming that the magnification is the same in all directions
($\sqrt{9.3} = 3.0$).

Though there is a fair amount of dispersion in the size of
individual galaxies in high-redshift dropout samples, the
mean size of galaxies at a given luminosity varies as $(1 +
z)^{-1.1 \pm 0.3}$ \citep[][see also \citealt{Ferguson2004,
Bouwens2004size}]{Bouwens2006}.  A1689-zD1 has a delensed \Hband\
magnitude of 27.1, which corresponds to $\sim 0.3 L^{*}_{z=3}$
\citep{Steidel1999}.  Scaling the mean sizes ($\sim$1.3~kpc) of
$\sim 0.3 L^{*}_{z=3}$ LBGs at $z \sim 3.8$ \citep{Bouwens2004size},
we would expect a mean half-light radius at $z \sim 7.6$ of $0.7
\pm 0.1$~kpc.  While this is slightly smaller than the half-light
radius of A1689-zD1, there is at least a $\pm 20\%$ dispersion in
galaxy sizes about the mean size at any given redshift.

\subsection{Counterimages}

We also used the improved \cite{Broadhurst2005} Abell~1689 strong
lensing model (see \S~\ref{sect:srcmag}) to look for potential
counterimages of A1689-zD1.  We re-projected the source image at
$z \sim 7.6$ back to the image plane at $z \sim 0.18$ and found
that the model does not predict any counterimages.  For a simple
isothermal sphere model with an Einstein radius of \RE, one would
expect that an object at radius $r$ would have a counterimage on
the opposite side of the cluster at a distance of $2 \RE - r$.
However, because of the complexity of the Abell~1689 lensing model,
not every background object is guaranteed to produce a counterimage.

\section{Ultraviolet (UV) Luminosity Function (LF)}

In theory our discovery of one very bright \zband-dropout in the
$\sim6$ arcmin$^{2}$ \Jband-band coverage around Abell~1689 should
allow us to estimate the volume density of UV-bright galaxies at
$z \sim 7$.  To explore this further, we generated mock images
based on the UV LF, sizes, and colors of LBGs at $z \sim 6$ found
by \cite{Bouwens2007} and \cite{Bouwens2006} and magnified them
according to the \cite{Broadhurst2005} Abell~1689 deflection map.
Unfortunately, from these no-evolution simulations we estimate that
we would expect to find only $\sim0.2$ galaxies over our search
area.  This suggests that there is very little that we could learn
about the UV LF from our \zband-dropout survey around Abell~1689,
and that, in fact, we may have been quite fortunate to find the
bright candidate A1689-zD1.  Even though the cluster magnification
increases the typical depth by $\sim2 - 3$ mag, the effective
source-plane area that is surveyed decreases inversely proportional
to the magnification factor.  The very small number of expected
candidates follows from the tiny area ($\la0.5$~arcmin$^{2}$) that
the \Jband-band coverage probes in the source plane at $z \ge 7$.

If the effective slope of the LF was large enough ($d(\log_{10}
N) / dm \ga 0.4$; \citealt{Broadhurst1995}), we would expect the
greater depth to more than make up for the smaller search area.
The only place the effective slope would be greater than $\sim$0.4
is for luminosities above \Lstar, unless we assume that the LF at
$z \sim 7-8$ is steeper than at $z \sim 6$ (i.e. $\alpha \approx
-1.74$, as determined by \citealt{Bouwens2007}, which corresponds to
$d(\log_{10} N) / dm \sim 0.3$).  Despite the small expected numbers,
the present search around Abell~1689 should be more efficient at
finding \Lstar\ galaxies at $z \ga 7$ than a search without the
lensing amplification.  However, lensing may not provide as great
an advantage in searches for galaxies fainter than \Lstar, given
recent measurements of the faint-end slope [where $d(\log_{10} N)
/ dm \sim 0.3$].

\section{Previous \lowercase{$z \ga 6$} Lensed LBG Candidates}

High-redshift $z \ga 6$ galaxy candidates have been previously
identified in searches around strong lensing clusters \citep{Hu2002,
Kneib2004, Richard2006}.  For example, in the field of the lensing
cluster Abell~2218, \cite{Kneib2004} found a very highly magnified
($\mu \sim 25$) and triply imaged LBG candidate near the critical
curve.  They infer a redshift for the candidate of $6.6 < z < 7.1$
from the broadband ACS and NICMOS photometry.  However, because of
the significant detection that their source shows in the \zband\ band
and the modest \zmJ\ $\sim 0.4$ break, we are confident that this
source is at a redshift lower than A1689-zD1 (where $\zmJ > 2.2$).

Using VLT ISAAC observations of Abell~1835 and AC114,
\cite{Richard2006} identified 13 (first and second category)
very bright LBG $z \ga 6$ candidates ($2.5\ \sigma$ detections),
some of which are $\sim 1$ mag brighter than A1689-zD1.  We can
evaluate the reliability of these sources by looking at the 5
candidates for which much deeper NICMOS \Hband-band data exist.
The deep NIC3 \Hband\ observations ($\sim 0.7$~arcmin$^{2}$)
of Abell~1835 and AC114 reach a $5\ \sigma$ limiting magnitude of
$\sim26.8$ AB, which is $\sim2$ mag deeper than their VLT ISSAC data.
Unfortunately, the 2 first and second category candidates (highest
confidence) and the 5 first, second, and third category candidates
(slightly lower confidence) covered by the NICMOS fields show
no detection ($< 2\ \sigma$) in the these data.  This suggests
that their sample of first, second, and third category $z \ga 6$
candidates is largely spurious \citep{Bouwens2008}.  Given its
secure detection significance ($8\ \sigma$), broadband colors,
and strength of its \zmJ\ break, A1689-zD1 is the brightest LBG
candidate that we can confidently place at a redshift $z \ga 7$.

\section{Summary and Conclusions}

We report the discovery of a strongly lensed LBG candidate
(A1689-zD1) at $z \sim 7.6$ in the field of the massive
galaxy cluster Abell~1689.  A1689-zD1 is $\sim$1.3 mag
brighter than the current brightest known \zband-dropout galaxy
\citep{Bouwens2006Nature, Bouwens2008}.  We find a strong \zmJ\ break
of at least 2.2 mag and best-fit photometric redshift of $z \sim
7.6$.  Employing a detailed cluster deflection model, we delensed
the source to examine its intrinsic properties.  We estimate a
magnification of $9.3$ at $z \sim 7.6$ at the position of A1689-zD1.
The high detection significance ($> 8\ \sigma$) of A1689-zD1 and
the consistency of the morphology in both the \Jband\ and \Hband\
bands provide strong evidence that the source is real and not a
spurious detection.  The source plane deprojection shows that the
star formation is occurring in compact knots of size $\la 300$~pc.

Using stellar population models to fit the rest-frame UV and optical
fluxes, we derive best-fit values for stellar masses \xten{(1.6 -
3.9)}{9} \Msun, stellar ages $45 - 320$~Myr, and star-formation
rates $< 7.6$~\Msunyr, properties generally similar to $z \sim 5 -
6$ LBGs.  A1689-zD1, with a redshift of $z \sim 7.6$ and a formation
redshift $z \ga 8.0$ ($t \la 0.63$ Gyr), is the brightest observed,
highly reliable $z > 7.0$ galaxy candidate found to date.

Given the unique brightness of A1689-zD1, we are actively working to
confirm its redshift with near-IR spectroscopy.  A spectrum of this
bright galaxy would allow for the most detailed study of a $z > 7$
LBG to date and provide valuable insights into the properties of
galaxies in the early universe.

\acknowledgments

We would like to thank Dan Coe, Andrew Zirm, and Brad Holden for
helpful discussions.  ACS was developed under NASA contract NAS
5-32865, and this research has been supported by NASA grants
NAG5-7697 and HST-GO10150.01-A and by an equipment grant from
Sun Microsystems, Inc.  The {Space Telescope Science Institute}
is operated by AURA Inc., under NASA contract NAS5-26555.  We are
grateful to K.~Anderson, J.~McCann, S.~Busching, A.~Framarini,
S.~Barkhouser, and T.~Allen for their invaluable contributions to
the ACS project at JHU.  L.I. acknowledges support from a Conicyt
Fondap grant.

%------------------------------------------------------------------------------
% Bibliography
%------------------------------------------------------------------------------

\end{document}